\documentclass[aps,prc,groupedaddress,showpacs,manuscript]{revtex4}
\usepackage{graphicx}

\usepackage{colordvi}
\usepackage{color}

\begin{document}

\title{The isospin dependent nucleon-nucleon inelastic cross section in the nuclear medium}

\author {Qingfeng Li$\, ^{1,2}$\footnote{E-mail address: liqf@zjhu.edu.cn} and
Zhuxia Li$\, ^{3}$}

\affiliation{
1) School of Science, Huzhou University, Huzhou 313000, P.R. China \\
2) Institute of Modern Physics, Chinese Academy of Sciences,  Lanzhou 730000,  P. R. China\\
3) China Institute of Atomic Energy, Beijing 102413, P. R. China\\
\\
 }
\date{\today}

\begin{abstract}

The calculation of the energy-, density-, and isospin-dependent $\Delta$ production cross sections in nucleon-nucleon (NN) scattering
$\sigma^*_{NN\rightarrow N\Delta}$ has been performed within the framework of the relativistic BUU approach. The N$\Delta$ cross sections are calculated in Born approximation taking into account the effective mass splitting of the nucleons and $\Delta$s in asymmetric matter. Due to the different mass splitting for neutron, proton and differently charged $\Delta$s, it is shown that, similar to the NN elastic ones, the reductions of N$\Delta$ inelastic cross sections in isospin-asymmetric nuclear medium are different from each other for all the individual channels and the effect is largest and of opposite sign for the $\Delta^{++}$ and $\Delta^{-}$ states. This approach is also compared to calculations without effective mass splitting and with splitting derived from Dirac-Brueckerner (DB) calculations. The isospin dependence of the N$\Delta$ cross sections is expected to influence the production of $\pi^+$ and $\pi^-$ mesons as well as their yield ratio, and thus affect the use of the latter quantity as a probe of the stiffness of the symmetry energy at supranormal densities.

\end{abstract}



\maketitle

\section{motivation}
Since the time when J. Chadwick proposed the existence of the neutron and soon afterwards W. Heisenberg introduced the new concept called isospin more
than 80 years ago, the picture of the nucleus consisting of
nucleons (neutrons and protons) has presented us its fascinating nature in a more understandable way, while
constantly producing new issues to be investigated. In nuclear
physics, the isospin dependent equation of state (EoS) of both
infinite and finite nuclear matter has been investigated in a
deeper and broader range recently \cite{Li:2008gp,Li:2014xx,Baldo:2016jhp}.
In addition to the isovector part of the mean field,
i.e., the symmetry potential
term, the isospin dependence
of the medium-modified cross sections of the collision term should also be considered
simultaneously
\cite{Mao2005aa,vanDalen:2005ns,Sammarruca:2005tk} for a real non-equilibrium process of heavy ion collisions, which is often simulated by microscopic transport models such as Boltzmann-Uehling-Uhlenbeck (BUU) or quantum
molecular dynamics (QMD).

A controversy concerning to the density dependence of the symmetry energy at supranormal densities has emerged in recent years.
Calculations with both the isospin-dependent
Boltzmann-Uehling-Uhlenbeck (IBUU) and the Lanzhou quantum
molecular dynamics (LQMD) models showed contradicting results
concerning the stiffness of the nuclear symmetry energy at
supranormal densities when they were compared with the same FOPI
experimental data for the $\pi^-/\pi^+$ yield ratios in heavy-ion reactions
\cite{Xiao:2008vm,Feng:2009am}. Since then, many more
calculations with, e.g., the ultrarelativistic quantum molecular
dynamics (UrQMD) model
\cite{Trautmann:2009kq,Russotto:2011hq,Wang:2014rva}, the
Boltzmann-Langevin approach (BL) \cite{Xie:2013np}, or the
T$\ddot{u}$bingen version of the QMD (TuQMD) \cite{Cozma:2013sja},
have entered into this topic but came up still with different conclusions. This situation even stirred the
re-examination of the technical model dependence. It involved 18 versions of
BUU- and QMD-type transport models and showed that sizable
uncertainties in nucleon-related observables exist
\cite{Xu:2016lue}.

The extraction of the stiffness
of the symmetry energy with the pion-related observables is considerably more demanding since pions are the decay products of resonances, mainly the
$\Delta(1232)$ at SIS energies. Hopefully the differences may be reduced after the detailed transport process of the $\Delta$
resonances in the nuclear medium has been more thoroughly investigated. Although
the modifications for resonances on the mean-field level
have been studied widely and often treated in models, the medium
modifications on $\Delta$-related collisions (more specifically,
the cross sections) are much less
thoroughly investigated in the transport model calculations. The situation is even more complicated regarding the isospin-asymmetric nuclear medium that
exists throughout the whole nonequilibrium process of a
heavy-ion collision (HIC). A first step on this topic was made using a covariant relativistic transport
approach \cite{Prassa:2007zw} by considering the density dependence of the nucleon-nucleon (NN) inelastic $NN\rightarrow
N\Delta$ cross section taken globally for all inelastic channels from a Dirac-Brueckner (DB) calculation. It is found that it appreciably
decreases the pion yield (owing to the reduction of the inelastic
cross section in the nuclear medium) but only moderately affects the
$\pi^-/\pi^+$ yield ratio.

In a more recent paper~\cite{Song:2015hua}, Song and Ko have considered this issue by investigating the effect of the isospin-dependent medium modification of the
pion-production threshold on the total pion yield and their ratio
with the help of the relativistic Vlasov-Uehling-Uhlenbeck (RVUU)
approach, based on mean fields from the nonlinear relativistic
NL$\rho$ and NL$\rho\delta$ models. It is found that, owing to the
reduced nucleon- and $\Delta$- masses in the medium, the threshold
effect obviously enhances the total yield. Owing to the
threshold difference for each pion-production channel, the
$\pi^-/\pi^+$ yield ratio is also enhanced, especially at lower
beam energies. But, the isospin effect on the cross sections of each of the $\Delta$ production channels is not taken into account.

In the current work, based on the effective Lagrangian within the
framework of the relativistic BUU (RBUU)
microscopic transport theory \cite{Mao2005aa,Li:2000sha,Li:2003vd}
in which the $\sigma$, $\omega$, $\delta[a_0(983)]$, $\rho$, and
$\pi$ mesons are coupled to both nucleons and $\Delta(1232)$
resonances, we focus on the medium modifications of NN inelastic
scatterings and especially in isospin-asymmetric circumstances. Note that the $\delta$ meson exchange is further taken into account on the basis of previous works as in Ref.~\cite{Li:2003vd} for the calculation of NN elastic cross sections.
Because of the splitting in both nucleon and $\Delta$
effective masses with the consideration of the $\delta$ meson exchange, the effective cross sections
$\sigma_{pp\rightarrow \Delta^{++}n}^*$ and
$\sigma^*_{nn\rightarrow \Delta^{-}p}$ are shown to be significantly
and inversely affected by the isospin asymmetry,
while the other cross sections, $\sigma^*_{pp\rightarrow \Delta^{+}p}$,
$\sigma^*_{pn\rightarrow \Delta^{+}n}$, $\sigma^*_{pn\rightarrow
\Delta^{0}p}$, and  $\sigma^*_{nn\rightarrow \Delta^{0}n}$ are
only weakly dependent on the isospin asymmetry. This will be discussed
in detail in the following sections.

The paper is arranged as follows. In the next section, the
RBUU microscopically based transport theory for both
nucleons and $\Delta(1232)$ resonances is introduced. The
mass splitting of isobars of
both particles in an isospin-asymmetric system and the effect on the NN
inelastic channels is emphasized. In sect. 3, numerical results for the energy-,
density-, and isospin- dependent cross section of the
$NN\rightarrow N\Delta$ process are shown and analyzed. Finally, a
summary and outlook is given in sect.4.

\section{Theoretical Framework}

The same framework is adopted as that in Refs.~\cite{Mao2005aa,Mao:1994zza}. So here we only provide the necessary formalism for the reader's convenience. By using the closed time-path Green's function technique (See, e.g., Refs.~\cite{Martin:1959jp,Schwinger:1960qe,Chou:1984es}, and the formalism was elaborated and developed further by Kadanoff and Baym and Keldysh \cite{Kadanoff1962,Keldysh1964}), the quantum transport equation, the so-called RBUU equation for describing the dynamic evolution of distribution functions of baryons, can be obtained after taking semi-classical and quasi-particle approximations, which reads

\begin{eqnarray}
&\{[\partial _x^\mu -\Sigma_{HF}^{\mu \nu }( x,p,\tau) \partial _\nu
^p-\partial _p^\nu \Sigma_{HF}^\mu(x,p,\tau) \partial _\nu ^x]p_\mu
+m^{*}_{N,\Delta}[\partial _\nu ^x\Sigma_{HF}^s( x,p,\tau) \partial _p^\nu  \nonumber \\
&-\partial _p^\nu \sum\nolimits_{HF}^s(x,p,\tau) \partial_\nu ^x]\}\frac{f_{N,\Delta}(x,p,\tau)}{E^{*}}=C(x,p,\tau) .
\label{eqrbuu}
\end{eqnarray}
The left side of Eq.~(\ref{eqrbuu}) is the mean-field part determined  by the mean-field self-energies. The right side is the collision part determined by the collisional self-energies and linked to the elastic and inelastic cross sections in medium, which will be discussed below. $m^{*}_N$ and $m_\Delta^*$ denote the
effective masses of nucleons and $\Delta$s, and $f_N$ and $f_\Delta$
are their single-particle distribution functions, respectively.

For the self-energies an approximation scheme has to be chosen. A natural choice is the Brueckner approximation, since it gives both the in-medium T-matrix and the mean field self energies consistently in the ladder approximation. This scheme has been developed in detail, e.g., in Refs.~\cite{Danielewicz:1982kk,TerHaar:1987ce}. It can be solved only in nuclear matter, but can be used in a transport calculation in the local density approximation. The Brueckner T-matrices and self energies are complicated functions of density and momentum, and are  difficult to use directly in a transport calculation. Thus approximation schemes have to be developed.

The mean field self energies have been parametrized in analogy with Quantum Hadrodynamics (QHD), where the self energy is given as the product of a meson-hadron coupling coefficient and the density corresponding to the spin-isospin character of the meson. This has been done, e.g., in Refs.~\cite{Fuchs:1995as,Brockmann:1990cn,Hofmann:2000vz}, where a momentum average is usually performed to make the coupling coefficients only density dependent. Note that this also means that no Fock terms should be included, since these are already taken into account in the Brueckner calculation. We also use, as in Refs.~\cite{Li:2003vd,Mao:1994np} before, this approximation here. This choice then also determines the effective masses, which also naturally become isospin dependent if the exchange of isovector $\rho$ and $\delta$ mesons are taken into account as in Ref.~\cite{Li:2003vd}.

Collisional self energies or in-medium effective cross sections are more difficult to parametrize, mainly because their momentum dependence is more complicated and important. It has been tried to directly use them in transport calculations \cite{Prassa:2007zw,Fuchs:1995as}. Rather than choosing the cross section empirically, as is most often done, we here follow, as in Refs.~\cite{Li:2003vd,Mao:1994zza,Chou:1984es} before, an intermediate way. We use the density-dependent mean-field meson-hadron coupling coefficients, and calculate cross sections in first-order Born approximation, using consistently the effective masses from the mean field calculation, and also including effective form factors. Although the coupling coefficients obtained from the mean-field level are not naturally proved to be valid for calculating cross sections which are beyond the mean field, it still seems to be a useful approximation as it was shown in Ref.~\cite{Mao:1994zza} that in this way one obtains reasonable results for elastic cross sections for zero and finite densities, as compared to Brueckner calculations and experiment, which in addition avoids any ambiguity related to an independent choice of the cross sections.

Therefore, in this work, the effective Lagrangian density for a system of nucleon and $\Delta$ baryons interacting though exchange of $\sigma$, $\omega$, $\rho$, $\delta$, and $\pi$ mesons is applied, which consists of the terms for free baryon and meson fields, $L_F$, and
the interaction part of baryons coupled to mesons, $L_I$.

The $L_F$ part reads as

\begin{eqnarray}
L_F^{} &=&\bar{\Psi}[i\gamma _\mu \partial ^\mu -m_N]\Psi +\bar{\Psi}_{\Delta\nu}[i\gamma _\mu \partial ^\mu -m_\Delta]\Psi_{\Delta}^{\nu} \nonumber \\
&&+\frac 12\partial
_\mu \sigma \partial ^\mu \sigma -\frac 14F_{\mu \nu }\cdot F^{\mu \nu }
+\frac {1}{2} \partial _{\mu} \mbox{\boldmath $\delta$}\partial ^{\mu}
\mbox
{\boldmath $\delta$}-\frac 14L_{\mu \nu }\cdot L^{\mu \nu }
+\frac {1}{2} \partial _{\mu} \mbox{\boldmath $\pi$}\partial ^{\mu}
\mbox
{\boldmath $\pi$} \nonumber \\
&&-U(\sigma )+U(\omega
)-U(\mbox{\boldmath $\delta$})+U(\mbox{\boldmath $\rho$})-U(\mbox{\boldmath $\pi$}),
\label{eq1}
\end{eqnarray}

where
\begin{equation}
F_{\mu \nu }\equiv \partial _\mu \omega _\nu -\partial _\nu \omega _\mu , \quad
L_{\mu \nu }\equiv \partial _\mu \mbox{\boldmath $\rho$}_\nu -\partial _\nu %
\mbox{\boldmath $\rho$}_\mu.
\label{eqfl}
\end{equation}
$U(\sigma )$, $U(\omega )$, $U(\mbox{\boldmath $\delta$} )$,
$U(\mbox{\boldmath $\rho$})$, and $U(\mbox{\boldmath $\pi$} )$  are the
self-interaction parts of the $\sigma $, $\omega $, $\delta $, $\rho $, and $\pi $  meson
fields and the respective expressions are

\begin{eqnarray}
&&U(\sigma )=\frac 12m_\sigma ^2\sigma ^2, \quad
U(\omega )=\frac 12m_\omega ^2\omega _\mu \omega ^\mu ,
\nonumber\\
&&U(\mbox{\boldmath $\delta$} )=\frac 12m_\delta ^2\mbox{\boldmath $\delta$}^2, \quad
U(\mbox{\boldmath $\rho$} )=\frac 12m_\rho ^2\mbox{\boldmath $\rho$}_\mu \mbox{\boldmath $\rho$
}^\mu , \nonumber\\
&&U(\mbox{\boldmath $\pi$} )=\frac 12m_\pi ^2\mbox{\boldmath $\pi$}^2.
\label{equsvdr}
\end{eqnarray}

The $L_I$ reads as
\begin{eqnarray}
L_I &=&g_{NN}^\sigma \bar{\Psi}\Psi \sigma -g_{NN}^\omega \bar{\Psi}\gamma _\mu \Psi
\omega ^\mu   +g_{NN}^\delta \bar{\Psi}\mbox{\boldmath $\tau$}\cdot \Psi
 \mbox{\boldmath $\delta$}-\frac 12g_{NN}^\rho \bar{\Psi}\gamma _\mu
\mbox{\boldmath $\tau$}\cdot \Psi \mbox{\boldmath $\rho$}^\mu \nonumber\\
&&+g_{\Delta\Delta}^\sigma \bar{\Psi}_{\Delta\nu}\Psi_{\Delta}^\nu \sigma -g_{\Delta\Delta}^\omega \bar{\Psi}_{\Delta\nu}\gamma _\mu \Psi_{\Delta}^\nu
\omega ^\mu   +g_{\Delta\Delta}^\delta \bar{\Psi}_{\Delta\nu}\mbox{\boldmath $\tau$}\cdot \Psi_{\Delta}^\nu
 \mbox{\boldmath $\delta$}-\frac 12g_{\Delta\Delta}^\rho \bar{\Psi}_{\Delta\nu}\gamma _\mu
\mbox{\boldmath $\tau$}\cdot \Psi_{\Delta}^\nu \mbox{\boldmath $\rho$}^\mu \nonumber\\
&&+g_{NN}^\pi \bar{\Psi}\gamma_\mu\gamma_5\mbox{\boldmath $\tau$}\cdot\Psi\partial^\mu\mbox{\boldmath $\pi$}
-g_{N\Delta}^\pi \bar{\Psi}_{\Delta\mu} \partial^\mu\mbox{\boldmath $\pi$}\cdot \mbox{\boldmath S}^+ \Psi
-g_{N\Delta}^\pi \bar{\Psi}\mbox{\boldmath S}\Psi_{\Delta\mu} \cdot \partial^\mu\mbox{\boldmath $\pi$}
\label{eq8}
\end{eqnarray}
where $\psi_{\Delta}$ is the Rarita-Schwinger spinor of the
$\Delta$ baryon which is same as in Refs.~\cite{Mao2005aa,Benmerrouche:1989uc}. $\tau$ is the isospin
operator of the nucleon, and $\mbox{\boldmath S}$ and
$\mbox{\boldmath S}^+$ are the isospin transition operators
between the isospin 1/2 and 3/2 fields. In principle, $\rho$ meson can also contribute to the $\Delta$ production but here we only take the N$\Delta\pi$ coupling into account since it is the most important production mechanism for $\Delta$.

For nucleons, the functional form of the density dependence of coupling as well as the parameters is taken from Ref.~\cite{Hofmann:2000vz}, which was deduced from DB calculations for describing isospin asymmetric matter and nuclei far from stability line and thus it is suitable for the aim of this study. As discussed above these couplings were also applied for the calculation of NN elastic cross sections in Ref.~\cite{Li:2003vd} and quite reasonable results were obtained. For the $\Delta$ isobars, the couplings to mesons are assumed to be the same as those of nucleons except the coupling to $\pi$, where $g_{NN}^\pi=f_\pi/m_\pi$
and $g_{N\Delta}^\pi=f_\pi^*/m_\pi$ are chosen and $f_\pi^2/4\pi=0.08$ and $f_\pi^{*2}/4\pi=0.37$, respectively \cite{Mao:1994zza}. The meson masses are $m_\sigma=550$
MeV, $m_\omega=783$ MeV, $m_\delta=983$ MeV, $m_\rho=770$ MeV, and
$m_\pi=138$ MeV, respectively.

\begin{figure}[htbp]
\centering
\includegraphics[angle=0,width=0.9\textwidth]{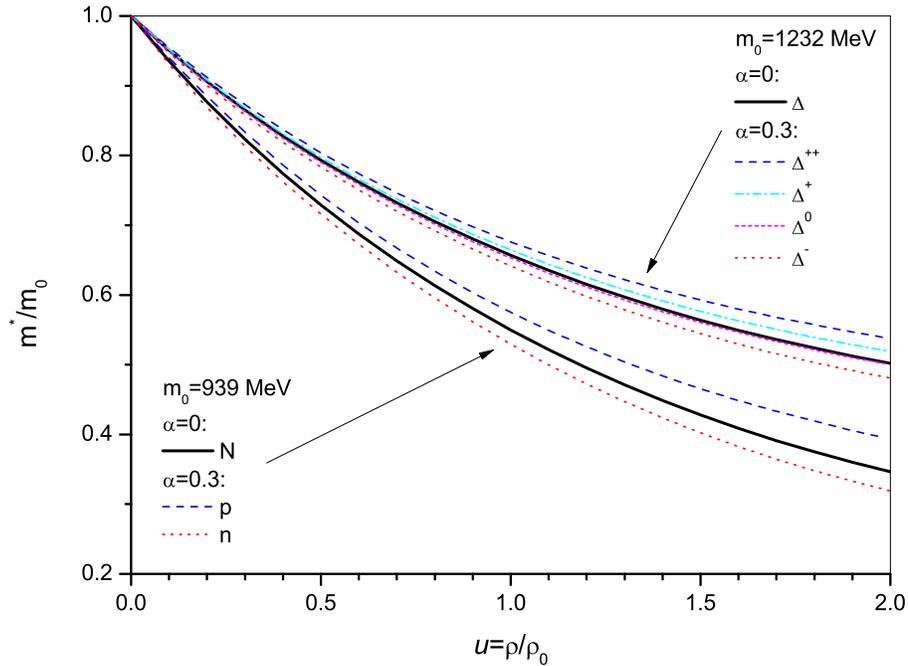}
\caption{\label{fig1} (Color online) The nucleon (lower group of
three lines) and $\Delta$ (upper group of five lines) effective
masses as a function of the reduced nuclear density. For each group,
calculations with the isospin asymmetry $\alpha=0$ and 0.3 are
shown. }
\end{figure}

Within the mean-field approximation and zero-temperature
assumption for particle distribution functions, the effective masses for protons, neutrons and four $\Delta$ isobars modified by the scaler part of mean-field read
\begin{eqnarray}
&&m_{p/n}^*=m_N-g_\sigma \sigma \mp g_\delta \delta_0 , \label{mpnstar} \\
&&m_{\Delta^{++}/\Delta^{-}}^*=m_\Delta-g_\sigma \sigma \mp g_\delta \delta_0 ,\label{mDGGMstar} \\
&&m_{\Delta^{+}/\Delta^{0}}^*=m_\Delta-g_\sigma \sigma \mp \frac13 g_\delta \delta_0 . \label{mDG0star}
\end{eqnarray}
Thus, as in Ref.~\cite{Song:2015hua} the effective masses of the $\Delta$ states are related to the nucleon effective masses by the Clebsch-Gordan coefficients of the $\Delta\rightarrow N\pi$ decomposition.
Note that the pion exchange has no contribution to the mean-field term since it is a pseudoscalar meson, and the value of
$\delta_0$ is negative in the neutron-rich nuclear medium. $m_{N}$ and $m_{\Delta}$ are the free nucleon ($m_N=939$ MeV) and
$\Delta$ masses (a resonance with the pole mass $m_\Delta=1232$ MeV
and a decay width about 120 MeV). Fig.~\ref{fig1} shows the density dependence of effective mass for protons, neutrons, $\Delta^{++}$, $\Delta^{+}$, $\Delta^{0}$, and $\Delta^{-}$ by solving the scaler mean-field and the meson field equations iteratively. The reduced effective masses $m^*/m_0$, where $m_0$ is the respective free mass,
vs. the reduced baryon density $u$ ($=\rho/\rho_0$) are calculated
with two values, 0 and 0.3, of the isospin asymmetry
$\alpha=(\rho_n-\rho_p)/\rho$. It has been found that, at the
reduced density $u$ less than about 2.5, the yield of $\Delta$s is
below 10$\%$ for energies below 1 GeV, as seen e.g. in Ref.~\cite{Li:2005gfa}. So, the
isospin asymmetry $\alpha$ using only densities of neutrons and
protons is accurate enough for the
current work and was adopted as well in the recent relativistic mean-field
calculations~\cite{Song:2015hua}. When $\alpha=0$ in the
isospin symmetric case, the splitting for both particles does not
appear since there is no contribution from the $\delta$-meson
field. Mainly because of the strong density dependence of the
$g_\sigma$ coupling constant, the decrease of $m^*/m_0$ with
density is obvious and thus certainly influences
the collision term (in-medium cross sections). This has been shown
for the NN elastic cross section in Ref.~\cite{Li:2003vd},
and will be further seen for the NN inelastic case. When $\alpha=0.3$ for the neutron-rich case,
the masses of $\Delta$-isobars differ and
$m^*_{\Delta^{++}}>m^*_{\Delta^{+}}>m^*_{\Delta^{0}}>m^*_{\Delta^{-}}$,
similar to the trend of nucleons. This mass splitting of
$\Delta$s  will definitely further influence the reaction channels
for the production of all $\Delta$-isobars, which will be
exhibited and discussed next.

Now, let us come to the calculation of the in-medium $\Delta$ production cross sections in the nucleon-nucleon scattering. The collision term of the RBUU equation can be divided into the elastic and inelastic parts. They can be further expressed in one form

\begin{eqnarray}
C_{el, in}(x,p,\tau)&=&\frac12\int{\frac{d^3p_2}{(2\pi)^3}\int\frac{d^3p_3}{(2\pi)^3}\int\frac{d^3p_4}{(2\pi)^3}(2\pi)^4 \delta^{(4)}(p_1+p_2-p_3-p_4)W_{el,in}(p_1,p_2,p_3,p_4)[F_2-F_1]}   \nonumber \\
&=&\frac12\int{\frac{d^3p_2}{(2\pi)^3}d\Omega\sigma^*_{el,in}(s,t,\alpha) v[F_2-F_1]} .
\label{eqcoll}
\end{eqnarray}
Here $W_{el,in}$ represents the transition probability and determines the elastic and inelastic differential cross sections $\sigma^*_{el}$ and $\sigma^*_{in}$ in medium, respectively. $F_1$ and $F_2$ are the Uehling-Uhlenbeck Pauli-blocking factors of the loss and gain terms, respectively. The four-momenta $p_1$ and $p_2$ are for ingoing particles while $p_3$ and $p_4$ for outgoing particles. The $v$ is the M$\phi$ller velocity, and variables $s$ and $t$ are Mandelstam
variables.

The transition probability for the inelastic reaction $NN\rightarrow \Delta N$ reads as
\begin{equation}
W_{in} (p_1,p_2,p_3,p_4)=G + G(p_3 \leftrightarrow p_4 \& \Delta \leftrightarrow N) ,
\label{win}
\end{equation}
where
\begin{equation}
G=\frac{(g_{NN}^\pi)^2(g_{N\Delta}^\pi)^2}{16 p_1^0 p_2^0 p_3^0 p_4^0} (T_d \Phi_d - T_e \Phi_e) .
\label{G1}
\end{equation}
Here $T_{d,e}$ and $\Phi_{d,e}$ are isospin and spin matrices for both direct and exchanged Feynman diagrams contributing to the lowest order collisional self-energy for $\Delta$ production (Born terms), which are given in Refs.~\cite{Mao2005aa,Mao:1994zza}. It is found that both
$T_d$ and $T_e$ are equal to 2 for $\Delta^{++}$ and $\Delta^{-}$
production channels, while they are $2/3$ for all other channels. $\Phi_{d}$ and $\Phi_{e}$ are the same in form with those in \cite{Mao2005aa,Mao:1994zza}, and are complicated functions of the momenta and masses of incoming and outgoing particles. However, it should be stressed that, due to the consideration of the isospin degree of freedom, the $\Phi_{d}$ and $\Phi_{e}$ functions for six channels of $\Delta$ production are actually different and have to be calculated {\it separately} when $\alpha \neq 0$.
And, with the help of the on-shell condition $p^2_{i,i=1-4}={m^*}^2_{i,i=1-4}$, the transition probability can then be solved analytically.

Before calculating the
inelastic cross sections, we consider effective form factors
for $NN\pi$ and $N\Delta\pi$ vertices. Commonly we choose the form $F_{NN\pi}(t)=\frac{\Lambda _{NN\pi}^2}{\Lambda _{NN\pi}^2-t}$
for the $NN\pi$ vertex, where $\Lambda_{NN\pi}$ is the cut-off mass for the exchanged pion
meson. For the $N\Delta\pi$ vertex, we noticed that there exist
various versions of the form factor partly due that the $\Delta$ is a decay particle \cite{Kloet:1984xi,TerHaar:1986xpv}. If we only consider
the behavior of the pole regardless of its mass distribution, as
will be focused upon in this paper, the effective form factor for the
$N\Delta\pi$ vertex will be same as that for $NN\pi$ and
$\Lambda_{NN\pi}=\Lambda_{N\Delta\pi}=510$ MeV as adopted in
\cite{Mao:1994np}.

Here, we should mention that we view the $\Delta$ particle as an elementary particle in the quasi-particle approximation. In fact, the mass distributions of the $\Delta$s are important especially at its production threshold, as shown in Refs.~\cite{Song:2015hua,Mao:1994zza}. However, the decay widths may depend on the states of the $\Delta$ isobar, which ought to be influenced by the nuclear medium as well, and is not clear at the moment. Further, the threshold effect, which was shown to be very important in Ref.~\cite{Song:2015hua}, and the screening of pion propagator, which was found to have a visible effect in Ref.~\cite{Mao:1994np}, are not taken into account as well. In this work, we want to focus on the problem of the isospin dependence of the $\Delta$ production cross sections in nucleon-nucleon scattering.  Therefore, although these problems mentioned above are certainly very important and should be considered carefully, their integrated effects will complicate the message of this work and will be studied in our next work.

\section{Numerical results}

\begin{figure}[htbp]
\centering
\includegraphics[angle=0,width=0.9\textwidth]{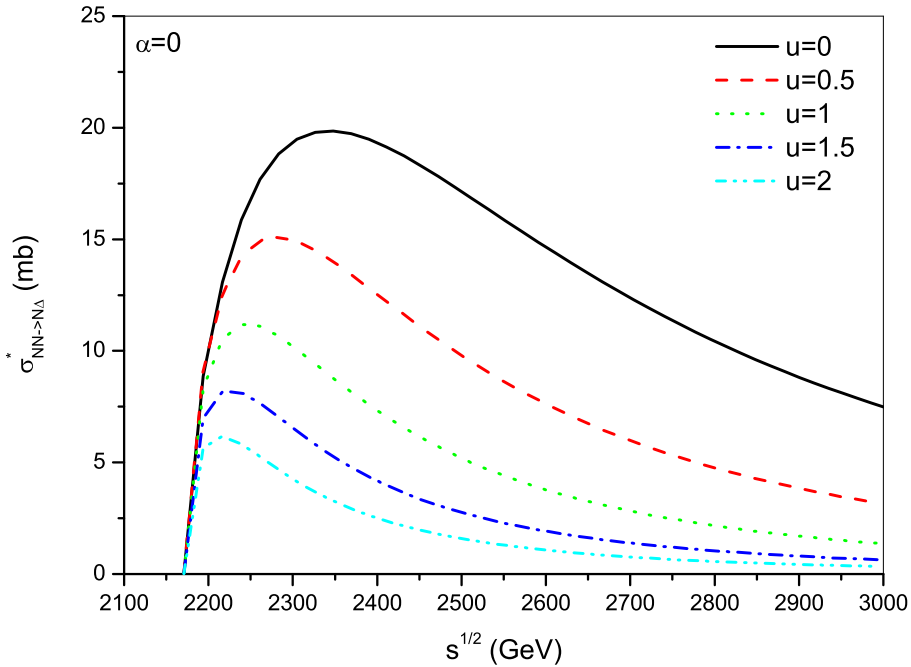}
\caption{\label{fig2} (Color online) The isospin-averaged in-medium cross section
$\sigma^*_{NN\rightarrow N\Delta}$ at several reduced densities
as a function of c.m. energies for symmetric nuclear matter ($\alpha=0$). }
\end{figure}

After integrating the $\Phi_{d}$ and $\Phi_{e}$ in Eq.~(\ref{G1}) over azimuth angle and averaging over the spin of initial states one can obtain the explicit expressions of all channels of $\Delta$ production \cite{Mao:1994zza}. Firstly, in Fig.~\ref{fig2} we show the in-medium cross section
$\sigma^*_{NN\rightarrow N\Delta}$ (the isospin averaged
one) at different reduced densities and center-of-mass (c.m.) energies. It is
found that at large energies, the $\sigma^*_{NN\rightarrow
N\Delta}$ monotonously decreases with increasing density. This is
similar to previous calculations (Ref.~\cite{Mao:1994np}) performed with
a different parameter set of the effective Lagrangian. But,
its density dependence is seen to be stronger than that in
Ref.~\cite{Mao:1994np}, a fact that is certainly mainly due to the strong
decrease of effective $\Delta$ and nucleon masses shown in
Fig.~\ref{fig1} and in Ref.~\cite{Li:2003vd}, respectively.   Further, when
approaching the threshold energy, the density dependence is
somewhat different from that in \cite{Mao:1994np} due to the
neglect of the $\Delta$ mass distribution, from which a further energy dependence arises both on the centroid $\Delta$ mass and on the effective form factor for the $N\Delta\pi$ vertex. This should play a more
important role at the lower energies.

If we quantitatively compare the current calculation when $u=0$ shown in Fig.~\ref{fig2} with the previous one in Ref.~\cite{Mao:1994np} and with the experimental data in Ref.~\cite{Shimizu:1982dx} (here a factor $3/4$ for the
isospin Clebsch-Gordan coefficient should be taken into account if the data for $\sigma_{pp\rightarrow
pp\pi^0+pn\pi^+}$ is in use) at a
typical kinetic energy $E_K=1$~GeV (correspondingly,
$s^{1/2}=2.326$~GeV) where the maximum of the cross section approaches, three values, 19.5 mb, 17.5 mb, and 18 mb are obtained, respectively, and obviously comparable well with each other. It is interesting to see that the current result at normal density ($u=1$) is also comparable to that of the DB calculations from Ref.~\cite{Mao:1994zza,TerHaar:1987ce}. E.g., at $E_K=1$~GeV, our result is about 10 mb, while the DB result is about 11-12 mb. This is understandable since the reduced effective mass $m^*/m_0$ at $u=1$ shown in Fig.~\ref{fig1} is about 0.55, while it was found in Ref.~\cite{Mao:1994zza} that the RBUU calculation where  $m^*/m_0=0.538$ can approach to the DB result where $m^*/m_0=0.605$.

\begin{figure}[htbp]
\centering
\includegraphics[angle=0,width=0.9\textwidth]{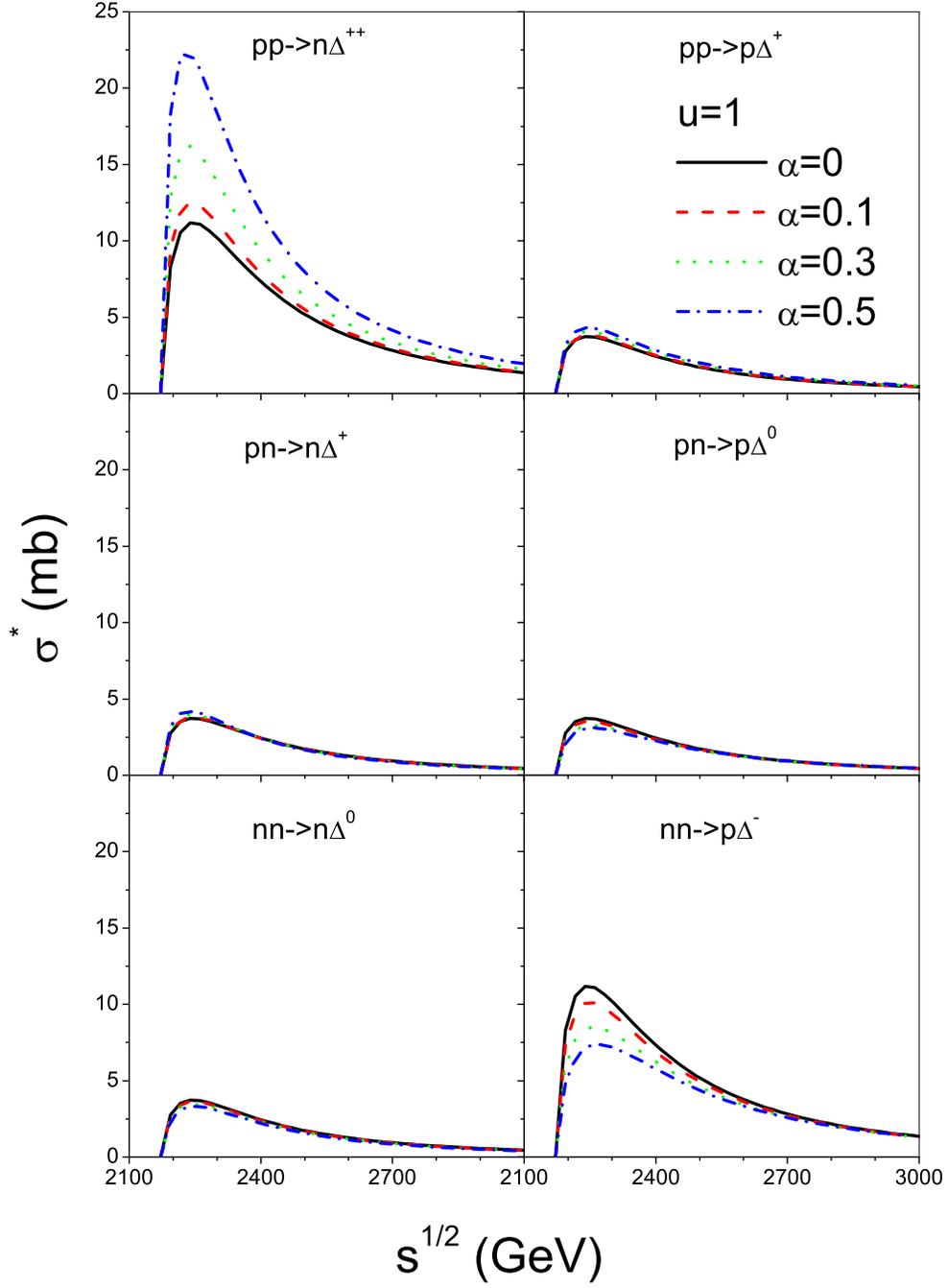}
\caption{\label{fig4} (Color online) The individual in-medium cross sections of the $NN\rightarrow N\Delta$ processes for $u=1$ and
for several $\alpha$ values as indicated with the different line styles.
}
\end{figure}

The results become more interesting for the isospin asymmetries
$\alpha \neq 0$ to be discussed in the following.
Fig.~\ref{fig4} depicts the energy dependence of individual cross
sections of the $NN\rightarrow N\Delta$ process. The $\alpha$
value varies from 0, 0.1, 0.3, up to 0.5 and the reduced density
$u=1$ is chosen. It was mentioned above that the isospin matrices $T_{d,e}$
in Eq.~(\ref{G1}) and consequently also
the cross sections for the production of $\Delta^{++}$
(top-left) and $\Delta^{-}$ (bottom-right) are exactly three times
larger than those of other channels as long as $\alpha=0$. With $\alpha \neq 0$,
this relation is lost, following from the effect of the mass splittings of
both nucleons and $\Delta$s in medium on the spin
matrices $\Phi_{d,e}$ in Eq.~(\ref{G1}). It is
found that the influence of the mass splitting on
$\sigma^*_{pp\rightarrow n\Delta^{++}}$ and
$\sigma^*_{nn\rightarrow p\Delta^{-}}$ is much stronger than that of the other
four channels, and the trends are opposite as $\alpha$ increases from 0
to 0.5, because the medium corrections of the effective
masses of $\Delta^{++}$ and $\Delta^{-}$ have opposite sign and
have larger values than those of $\Delta^{+}$ and
$\Delta^{0}$ when increasing $\alpha$ [see Eqs.~(\ref{mpnstar})-(\ref{mDG0star})]. Further, it is seen that, regardless of the accompanying production of either the neutron or the proton, the trend for the production of $\Delta^{+}$ ($\Delta^{0}$) follows that for $\Delta^{++}$ ($\Delta^{-}$) since the splitting
effect of the $\Delta$ masses on $\Phi_{d,e}$ is larger than that of the nucleon masses.

\begin{figure}[htbp]
\centering
\includegraphics[angle=0,width=0.9\textwidth]{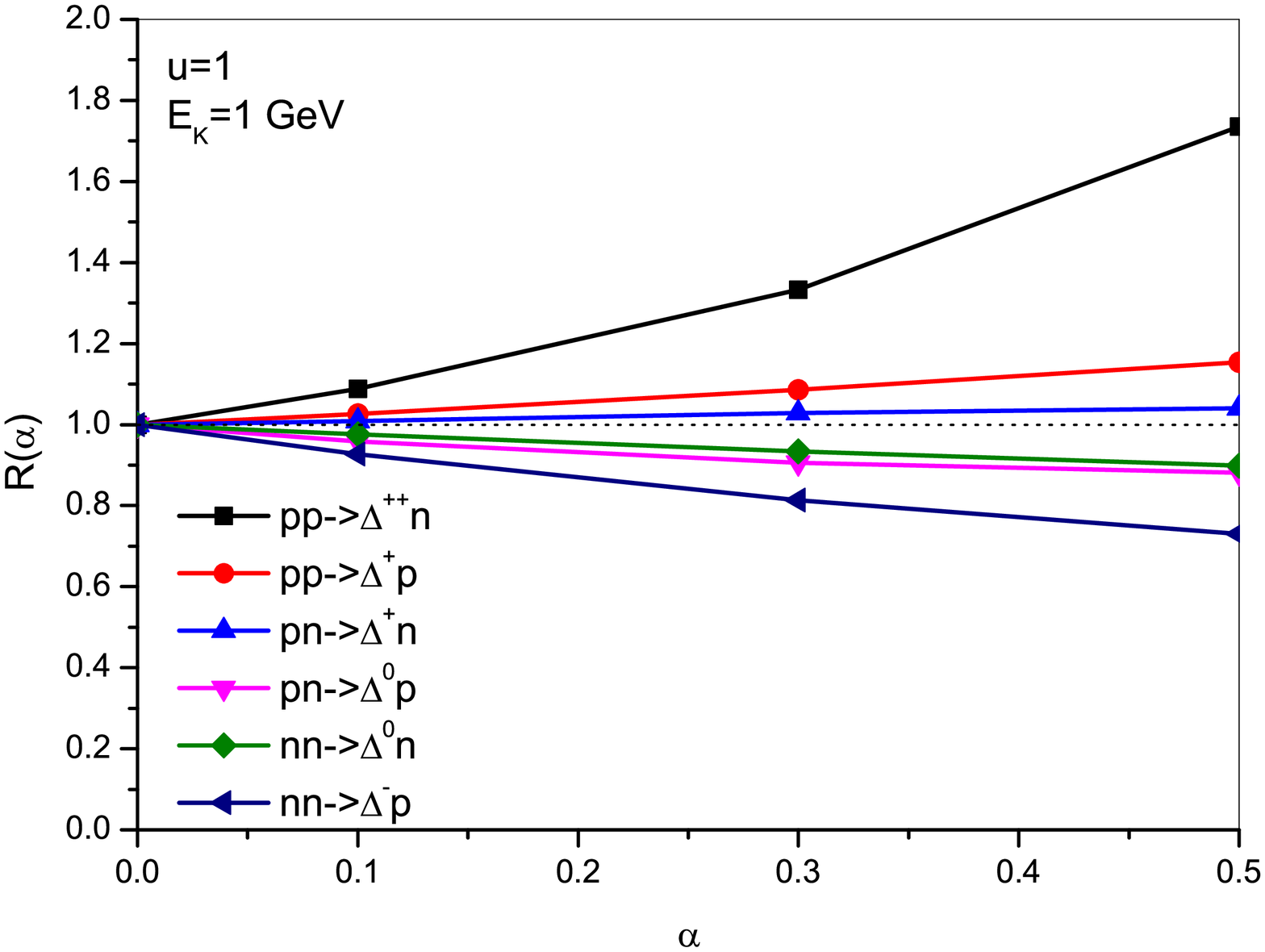}
\caption{\label{fig5} (Color online) The $R(\alpha)=\sigma^* (\alpha)/\sigma^* (\alpha=0)$ ratios of all channels (lines with different symbols)
as a function of the isospin asymmetry for $u=1$ and $E_K=1$~GeV. The horizontal dotted line represents unity.
}
\end{figure}

To see the effect of the mass splitting more clearly, the ratio
$R(\alpha) =\sigma^* (\alpha)/\sigma^* (\alpha=0)$ of all
channels is shown as an example in  Fig.~\ref{fig5}
as a function of the isospin asymmetry for $E_K=1$ GeV.
The ratio $R(\alpha)$
deviates almost linearly from unity when the value of the isospin
asymmetry increases from 0 to 0.5. It occurs with the sequence:
$R(\alpha,pp\rightarrow n\Delta^{++})
>R(\alpha,pp\rightarrow p\Delta^{+}) > R(\alpha,pn\rightarrow
n\Delta^{+}) > R(\alpha,pn\rightarrow p\Delta^{0}) > R(\alpha,
nn\rightarrow n\Delta^{0}) > R(\alpha,nn\rightarrow
p\Delta^{-})$. It is further seen that, at $\alpha=0.5$, the $R(\alpha)$ ratio remains within the interval
between 0.88 and 1.15 for the $\Delta^{+}$ and
$\Delta^{0}$ production channels, while it changes more rapidly to 1.74 and 0.73 for $\Delta^{++}$ and $\Delta^{-}$, respectively.

It is noted here that the isospin influenced NN (in-)elastic cross
sections in medium have been employed in calculating pion
production from neutron-rich HICs at the beam energy 400
MeV$/$nucleon \cite{Yong:2010zg}, and a visible effect on the
pion yield as well as the $\pi^-/\pi^+$ ratio has been observed.
However, the trend of the mass splitting for whatever nucleons or
$\Delta$s is quite different and even controversial, which has attracted continuous attention \cite{vanDalen:2005ns,Long:2005ne,Xu:2010fh,Zhang:2014sva,Guo:2015aa} and, certainly, leads to different isospin dependent cross sections in medium. For
example, in  \cite{Yong:2010zg} the $\sigma^* (\alpha>0)/\sigma^*
(\alpha=0)$ ratio follows the trend of $nn > np > pp$ of ingoing
colliding pairs without distinguishing the final outgoing states.
In our current calculations, the final states are also
considered and a contrary trend is seen for the ratio of cross sections as stated above. As a consequence, more $\pi^+$s would be produced with our cross sections than $\pi^-$s from the neutron-rich colliding system so as to drive down the $\pi^-/\pi^+$ ratio. We notice that the trend of the mass splitting for both  nucleons and $\Delta$s is the same as that in Ref.~\cite{Song:2015hua}, where an enhancement of the $\pi^-/\pi^+$ ratio is seen in Au+Au collisions. Also, the large cancellation of the isospin effects from both the $\Delta$ production threshold and the NN inelastic cross section deserves much attention as well.
In any case, the newly observed large
isospin effect on the production of $\Delta^{++}$ and $\Delta^{-}$
resonances is worth studying in microscopic transport model calculations for
real HICs.

\section{Summary and Outlook}

In the current work, theoretical calculations on the energy-, density-, and $\emph{isospin}$-dependent nucleon-nucleon (NN)
inelastic cross sections $\sigma^*_{NN\rightarrow N\Delta}(s,\rho,\alpha)$ are accomplished with the help of the
RBUU microscopic transport theory in which the $\sigma$, $\omega$, $\delta[a_0(980)]$, $\rho$, and $\pi$ mesons are coupled by density-dependent coefficients
to both nucleons and $\Delta(1232)$ resonances. Similar to previous isospin-averaged calculations, the decrease of
$\sigma^*_{NN\rightarrow N\Delta}$ with the increase of density is seen but relatively stronger due to a different choice of
the density-dependent parameter set for the effective Lagrangian. Due to the mass-splitting effect of both nucleons and $\Delta$s with the consideration of the $\delta$ meson exchange, all individual channels of the $\sigma^*_{NN\rightarrow N\Delta}$
are now also different from each other in the isospin-asymmetric nuclear medium with trends similar to that of the NN elastic
cross section $\sigma^*_{NN\rightarrow NN}$.
In addition, the largest but opposite in sign isospin effect is seen in the production of $\Delta^{++}$ and $\Delta^{-}$ resonances. In calculations
for heavy-ion collisions, it will influence the production of $\pi^+$ and $\pi^-$ mesons, as well as their yield ratio.

As a next step in future work, the medium (especially the isospin-dependent) modifications in the mean field, the production threshold,
and the mass distribution of the $\Delta$ resonance will be considered, together with the modifications of cross sections in
numerical calculations with a microscopic transport model.
The influence on the pion production in HICs at SIS energies including the strong principle of detailed balance for $\sigma^*_{NN\rightarrow N\Delta}$
and the mass splitting of both nucleons and $\Delta$s will be examined more thoroughly and presented in a future publication. This will hopefully lead to more reliable conclusions on the high-density symmetry energy.

\begin{acknowledgments}
We wish to thank H. St$\ddot{o}$cker, M. Bleicher, S. Schramm, T. Song, and W. Trautmann for valuable discussions and acknowledge support by the computing servers C3S2 in Huzhou
University and the LOEWE-CSC in FIAS institute of Frankfurt University. The work is financially supported in part by the National
Natural Science Foundation of China (Nos. 11375062, 11547312, 11275052, 11475262, 11475004, 11647306). Q.L. acknowledges the warm hospitality of
FIAS institute during the stay in the middle of the year 2016.
\end{acknowledgments}

\end{document}